%% file: Formatting-Instructions-LaTeX-2024.tex
\documentclass[letterpaper]{article} 
\usepackage{aaai24}  
\usepackage{times}  
\usepackage{helvet}  
\usepackage{courier}  
\usepackage[hyphens]{url}  
\usepackage{graphicx} 
\usepackage{natbib}  
\usepackage{caption} 
\usepackage{algorithm}
\usepackage{algorithmic}

\usepackage{newfloat}
\usepackage{listings}
\DeclareCaptionStyle{ruled}{labelfont=normalfont,labelsep=colon,strut=off} 
\floatstyle{ruled}
\newfloat{listing}{tb}{lst}{}
\floatname{listing}{Listing}
\usepackage{times}
\usepackage{epsfig}
\usepackage{graphicx}
\usepackage{amsmath}
\usepackage{amssymb}

\usepackage{tikz}
\usepackage{float}
\usepackage{subcaption}
\usepackage{booktabs}
\usepackage{hhline}
\usepackage{multirow}
\usepackage{tabularx}
\usepackage{filecontents}
\usepackage{caption}
\usepackage{xspace}

\usepackage{pifont}

\DeclareMathOperator*{\argmin}{arg\,min}

\newcommand{\tool}{{\sc \texttt{Deep3DMark}}\xspace}
\newcommand{\jiang}{{\sc \texttt{Jiang2017}}\xspace}
\newcommand{\spectral}{{\sc \texttt{Spectral}}\xspace}
\newcommand{\peng}{{\sc \texttt{Peng2021}}\xspace}
\newcommand{\ppeng}{{\sc \texttt{Peng2022}}\xspace}
\newcommand{\tsai}{{\sc \texttt{Tsai2020}}\xspace}
\newcommand{\ttsai}{{\sc \texttt{Tsai2022}}\xspace}
\newcommand{\hou}{{\sc \texttt{Hou2023}}\xspace}
\newcommand{\wang}{{\sc \texttt{Wang2022}}\xspace}
\newcommand{\yoo}{{\sc \texttt{Yoo2022}}\xspace}

\title{Rethinking Mesh Watermark: Towards Highly Robust and Adaptable Deep 3D Mesh Watermarking}
\author {
    Xingyu Zhu\textsuperscript{\rm 1,\rm 2,\rm 3},
    Guanhui Ye\textsuperscript{\rm 2},
    Xiapu Luo\textsuperscript{\rm 3},
    Xuetao Wei\textsuperscript{\rm 2,\rm 1}\thanks{Corresponding Author.}
}
\affiliations {
    \textsuperscript{\rm 1}Research Institute of Trustworthy Autonomous Systems, \\ Southern University of Science and Technology, Shenzhen 518055, China \\
    \textsuperscript{\rm 2}Department of Computer Science and Engineering, Southern University of Science and Technology, China \\
    \textsuperscript{\rm 3}Department of Computing, Hong Kong Polytechnic University, Hong Kong\\
    12150086@mail.sustech.edu.cn, 12132370@mail.sustech.edu.cn, csxluo@comp.polyu.edu.hk, weixt@sustech.edu.cn
}

\usepackage{bibentry}

\begin{document}

\maketitle

\input{chapter/abstract}

\input{chapter/introduction}
\input{chapter/related}
\input{chapter/method}
\input{chapter/experiment}

\section*{Acknowledgment}
This work was supported in part by National Key R\&D Program of China under Grant 2021YFF0900300, in part by Key Talent Programs of Guangdong Province under Grant 2021QN02X166, and in part by Research Institute of Trustworthy Autonomous Systems under Grant C211153201. Any opinions, findings, and conclusions or recommendations expressed in this material are those of the author(s) and do not necessarily reflect the views of the funding parties.

\bibliography{citation/3Dlearner, citation/adaption, citation/experiment, citation/meshwm, citation/others}

\end{document}


\maketitle

\input{chapter/supplementary}

\bibliography{citation/3Dlearner, citation/adaption, citation/experiment, citation/meshwm, citation/others}

%% file: chapter/abstract.tex
\begin{abstract}

The goal of 3D mesh watermarking is to embed the message in 3D meshes that can withstand various attacks imperceptibly and reconstruct the message accurately from watermarked meshes. The watermarking algorithm is supposed to withstand multiple attacks, and the complexity should not grow significantly with the mesh size. Unfortunately, previous methods are less robust against attacks and lack of adaptability. In this paper, we propose a robust and adaptable deep 3D mesh watermarking \tool that leverages attention-based convolutions in watermarking tasks to embed binary messages in vertex distributions without texture assistance. Furthermore, our \tool exploits the property that simplified meshes inherit similar relations from the original ones, where the relation is the offset vector directed from one vertex to its neighbor. By doing so, our method can be trained on simplified meshes but remains effective on large size meshes (size adaptable) and unseen categories of meshes (geometry adaptable). Extensive experiments demonstrate our method remains efficient and effective even if the mesh size is 190$\times$ increased. Under mesh attacks, \tool achieves 10\%$\sim$50\% higher accuracy than traditional methods, and 2$\times$ higher SNR and 8\% higher accuracy than previous DNN-based methods.
\end{abstract}

%% file: chapter/introduction.tex
\section{Introduction}

Digital watermarking is a technology used in copyright protection of multimedia, such as images, videos, point clouds, and meshes. The goal of digital watermarking is to obtain watermarked media by embedding messages in the media in the embedding phase and reconstructing the message from the watermarked media in the reconstruction phase. However, previous 3D mesh watermarking methods pursue high capacity while ignoring robustness. The watermark should be \textit{imperceptible} and \textit{robust} so that it can withstand attacks and be \textit{adaptable} so that it can be applied to arbitrary mesh sizes and geometries.


\begin{figure}[t]
    \centering
    \includegraphics[width=\linewidth]{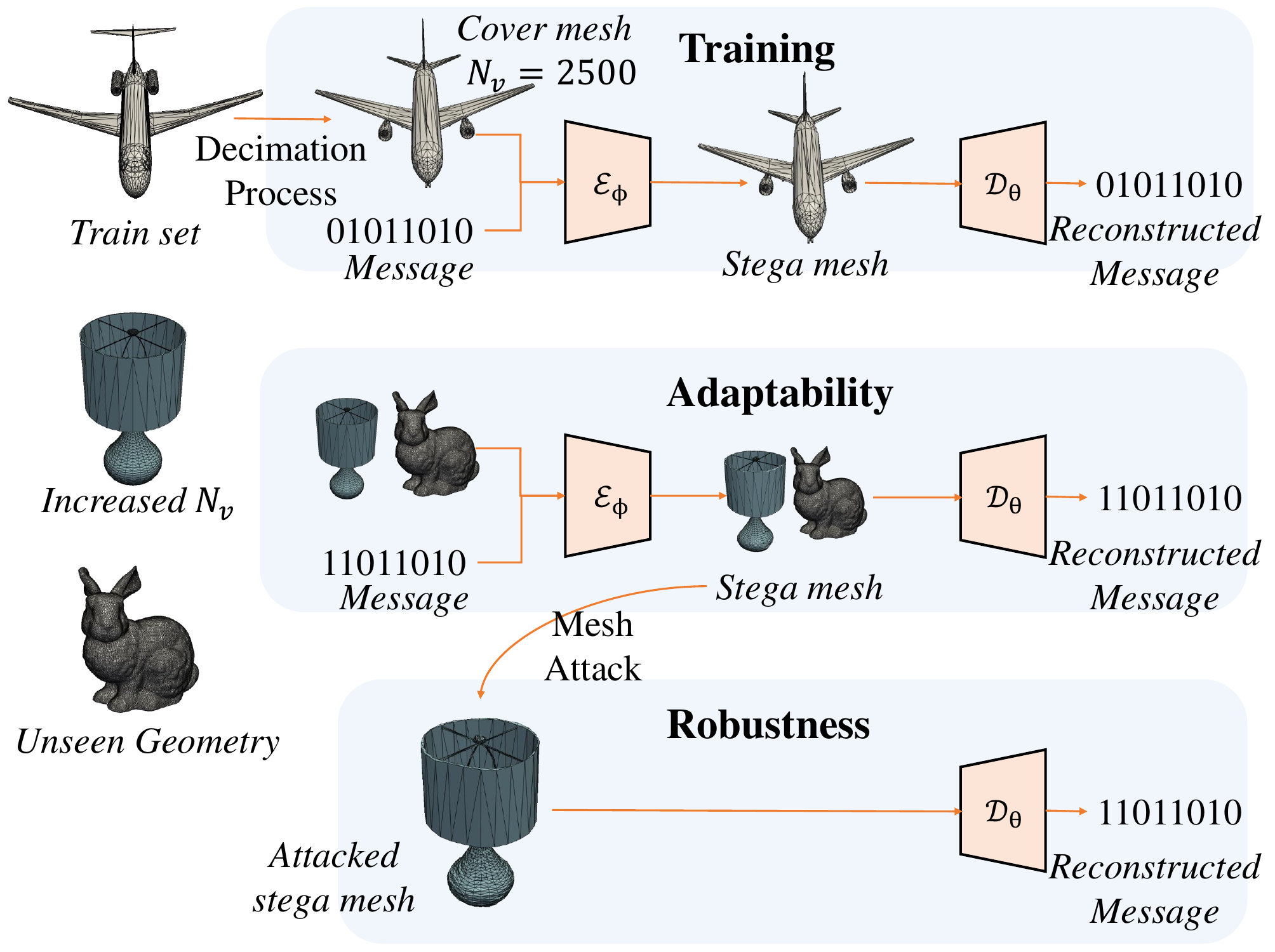}
    \caption{We train our \tool on simplified meshes (\textbf{Top}) and test on varied mesh sizes and unseen geometry (\textbf{Middle}) to show adaptation. We further test \tool on multiple mesh attacks (\textbf{Bottom}).}
    \label{fig:what_we_are_doing}
\end{figure}

\begin{figure*}[htb]
    \centering
    \includegraphics[width=1.0\textwidth]{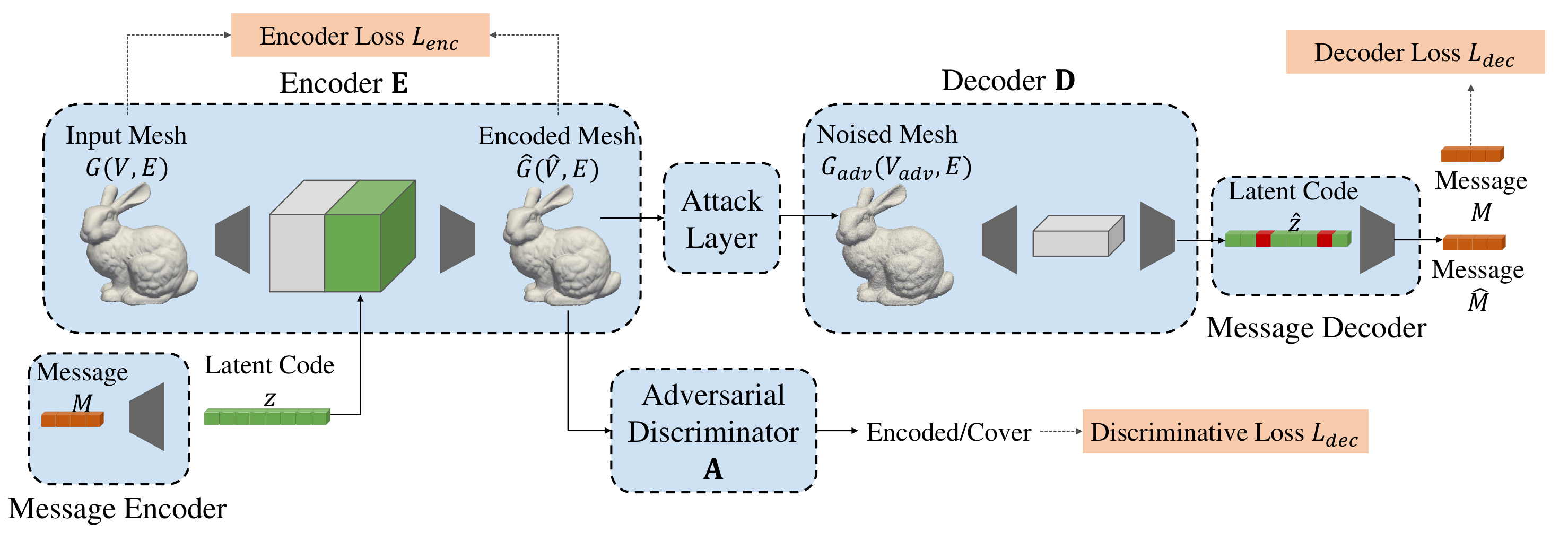}
    \caption{\tool overview. Message encoder first maps message $M$ into latent code $z$, which is further fed to the watermark encoder $\mathbf{E}$ along with the input mesh $G$ to generate the encoded mesh $\hat{G}$. The attack layer generates a noised mesh $G_{adv}$. Given the noised mesh, the watermark decoder $\mathbf{D}$ produces the decoded latent code $\hat{z}$ followed by the message decoder, which decodes latent code into decoded message $\hat{M}$. The adversarial discriminator encourages minimizing the difference between $G$ and $\hat{G}$.}
    \label{fig:arch}
\end{figure*}

Previous 3D mesh watermarking methods can be classified into DNN-based and traditional methods. Traditional methods focus on improving the capacity of watermarking (\ie, the number of embedded bits per vertices) while ignoring the robustness of their watermark. For example, some methods \cite{peng2021general, tsai2020separable} embed secret messages in the least significant bits (LSBs) of vertex coordinates, but they are vulnerable to Gaussian noises. Others \cite{tsai2022integrating, hou2023separable} embed secret messages in the most significant bits (MSBs) of vertex coordinates, which makes them robust against noises. However, they still cannot withstand rotation and scaling. Recent DNN-based methods show the possibility of embedding watermarks in either vertex domain \cite{wang2022deep} or texture domain \cite{yoo2022deep}. However, these methods are either only able to extract messages with textured meshes\cite{yoo2022deep} or low in watermarked mesh quality \cite{wang2022deep}. Moreover, watermark quality and embedding overhead should be less impacted by variations in mesh sizes and geometries for practical application consideration. Previous work has yet to explore a robust and adaptable watermarking method.

In this paper, we propose a highly robust and adaptable deep 3D watermarking \tool. Compared to traditional methods, \tool is more robust against multiple mesh attacks even without prior knowledge of the type of attacks. Compared to previous DNN-based methods, \tool generates higher-quality watermarked meshes and can be applied to different mesh sizes and unseen geometries. We watermark the vertex coordinates of meshes by adopting graph-attention network (GAT) \cite{velivckovic2017gat}. To achieve robustness, we apply adversarial training with GAT-generated perturbations. To achieve adaptability, we provide an insight that different sizes of meshes under the same categories share similar features. By exploiting such similarities, we can train our \tool with simplified meshes and keep effective on meshes with increased sizes.

Specifically, our adopted GAT is the backbone of our \tool, which can generate watermarked meshes given the original meshes and binary messages and reconstruct the binary messages from the watermarked meshes. Our \tool consists of 1) an encoder, 2) a decoder, 3) a message autoencoder, 4) an attack layer, and 5) a discriminator, as shown in Fig. \ref{fig:arch}. We train our \tool using simplified data from the train set under all scenarios to better evaluate the adaptability. To prove effectiveness, our \tool is tested on complete test data. To prove robustness, we test \tool on multiple mesh attacks. To prove adaptability, we test \tool on multiple datasets and different sizes of meshes. In summary, our contributions are the following:
\begin{itemize}
    \item We investigate mainstream watermarking methods and observe their low robustness. To tackle this problem, we propose a highly robust deep 3D watermarking \tool, which embeds binary messages in vertex distributions by incorporating the graph attention network (GAT) and achieves robustness against unknown mesh attacks by incorporating adversarial training.
    \item We achieve adaptability by exploiting the property that simplified meshes inherit similar relations from the original meshes, where the relation is an offset vector directed from a vertex to its neighbor. Our \tool can be trained on simplified meshes but remains effective on large-sized meshes and unseen categories of meshes.
    \item We conduct extensive experiments on various datasets to prove \tool's effectiveness, robustness, and adaptability. Our \tool achieves 10\%$\sim$50\% higher accuracy when facing attacks compared to traditional methods and achieves 50\% lower distortions and 8\% higher accuracy compared to previous DNN-based methods. Our experiment shows that our method can also be robust against multiple unknown attacks.
\end{itemize}


%% file: chapter/related.tex
\section{Related Work}
\begin{figure*}[htb]
    \centering
    \includegraphics[width=0.8\linewidth]{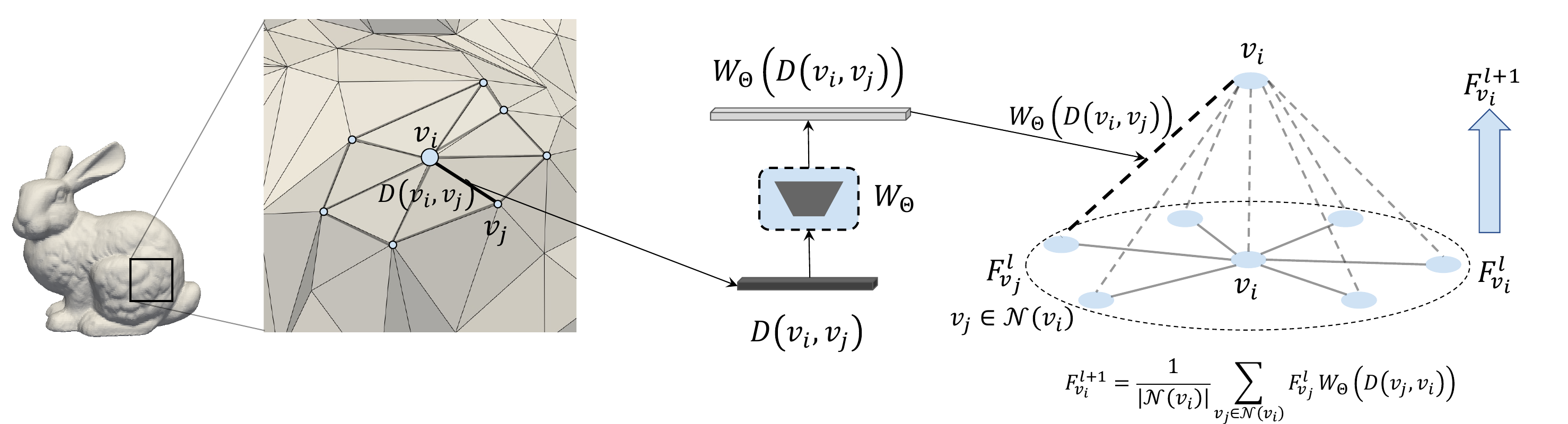}
    \caption{The process of generating the new feature $F^{l+1}_{v_i}$ of $v_i$. \textbf{Left:} the local region centered at $v_i$. \textbf{Middle:} within neighborhood $v_j\in \mathcal{N}(v_i)$, an MLP generates weights between $v_i$ and $v_j$ given their relation $D(v_i, v_j)$. \textbf{Right:} given vertex feature $\mathbf{F}^l=\{F^l_{v_0}, F^l_{v_2}, ..., F^l_{v_N}\}$ and generated weights of neighborhood $v_j\in \mathcal{N}(v_i)$, compute the new feature vector $F^{l+1}_{v_i}$ as the weighted sum of neighbor features.}
    \label{fig:GAT}
\end{figure*}

\subsection{Mesh Watermarking}

Early 3D mesh watermarking methods \cite{son2017perceptual, al2019graph} used Fourier and wavelet analysis to transfer meshes into the frequency domain and embed watermark bits into Fourier/wavelet coefficients. However, the time complexity of these methods grows cubically with the number of vertices. \cite{zhou2018distortion, jiang2017reversible, tsai2022integrating, hou2023separable} proposed to embed watermarks into the least significant bits and the most significant bits of vertex coordinates. \cite{hou2017blind} leveraged the layering artifacts of 3D printed meshes to apply watermark embedding and reconstruction.

Recently, \cite{yoo2022deep} and \cite{wang2022deep} explored the feasibility of DNN in watermarking work. \cite{wang2022deep} stacked graph residual blocks to embed and extract the watermark. \cite{yoo2022deep} embedded secret messages in textures of meshes and then extracted the message from the rendered 2D image, but cannot reconstruct an accurate message without the help of a texture encoder. In this case, replacing the texture image can completely remove the watermarks. Moreover, recent research \cite{dong2022think} also shows how to detect DNN-generated images. Hence, watermarking in mesh geometries is more secure than watermarking in textures.

\subsection{Neural Networks for 3D Meshes}

Existing methods for 3D data built features from faces \cite{xu2017dcn, feng2019meshnet, lian2019meshsnet, hertz2020DGTS, hu2022subdivision, kim2022exmeshcnn}, edges \cite{simonovsky2017ecc, velivckovic2017gat, wang2019dgcnn} and vertices \cite{qi2017pointnet, qi2017pointnet++, wu2019pointconv, liu2019rscnn, xu2018spidercnn, hermosilla2018montecnn, groh2018flex}. The built features were applied to downstream tasks such as classification and segmentation. \cite{velivckovic2017gat, simonovsky2017ecc} introduced an attention-based mechanism into graph convolution, where the weights of each neighbor were adjusted based on the edge information. Such graph-based convolution can be further extended to 3D meshes.



%% file: chapter/method.tex
\section{Definition} \label{sec:def}

Triangle meshes can be viewed as undirected graphs $G(V,E)$. Vertices $V\in \mathbb{R}^{N_v\times C_v}$ contains $N_v$ vertices, and each vertex has $C_v$ vertex elements such as coordinates and normals. Edges $E$ can be transformed from faces set of triangle meshes, where each face is a triangle formed by three vertex indices. Since changes of $E$ produce unexpected artifacts, we embed a binary message $M\in\{0,1\}^{N_m}$ into the vertex distribution $V$, i.e., we embed binary messages in vertex distributions $V$. Let $V,\hat{V},M,\hat{M}$ denote the original vertex, watermarked vertex, binary messages, and reconstructed messages, respectively. We model the problem by the following equations:
\begin{equation}
    \label{eq:problem}
    \begin{gathered}
        \hat{V} = \mathcal{E}_\phi(V, M)\\
        \hat{M} = \mathcal{D}_\theta(\hat{V})
    \end{gathered}
\end{equation}
In Eq \ref{eq:problem}, a parameterized encoding function $\mathcal{E}_\phi$ generates watermarked vertices $\hat{V}$ given original vertices $V$ and a binary message $M$. A parameterized decoding function $\mathcal{D}_\theta$ reconstructs $\hat{M}$ from $\hat{V}$. The encoding function should minimize the perturbation between $V$ and $\hat{V}$ by minimizing the following loss to achieve imperceptible embedding:
\begin{equation}
    \label{eq:obj1}
    \begin{aligned}
        &L_{enc}(\phi, \theta)=\mathbb{E}_{V,M}[\lVert \hat{V} - V \rVert_2^2]
    \end{aligned}
\end{equation}
To achieve precise reconstruction, we try to minimize the following loss:
\begin{equation}
    \label{eq:obj2}
    \begin{aligned}
        &L_{dec}(\phi, \theta)=\mathbb{E}_{V,M}[\lVert \hat{M} - M \rVert_2^2]
    \end{aligned}
\end{equation}
Finally, we have combined the optimization problem:
\begin{equation}
    \label{eq:finalobj}
    \phi^{*}, \theta^{*} = \argmin_{\phi, \theta} (L_{enc}(\phi, \theta) + L_{dec}(\phi, \theta))
\end{equation}


\section{Method} \label{sec:GAN}

We propose \tool, an end-to-end imperceptible watermarking method that can be robust to arbitrary attacks and be adaptable to different mesh sizes and geometries. To watermark a graph signal $G(V, E)$, we utilize local features in the spatial domain using graph attention network (GAT), which is the backbone of our \tool. We first introduce GAT on mesh. Then we introduce all \tool modules, followed by a detailed introduction to our training details.

\subsection{Graph Attention Network on Mesh} \label{sec:gat}

\begin{figure}
    \centering
    \begin{subfigure}[b]{0.49\linewidth}
        \includegraphics[width=\linewidth]{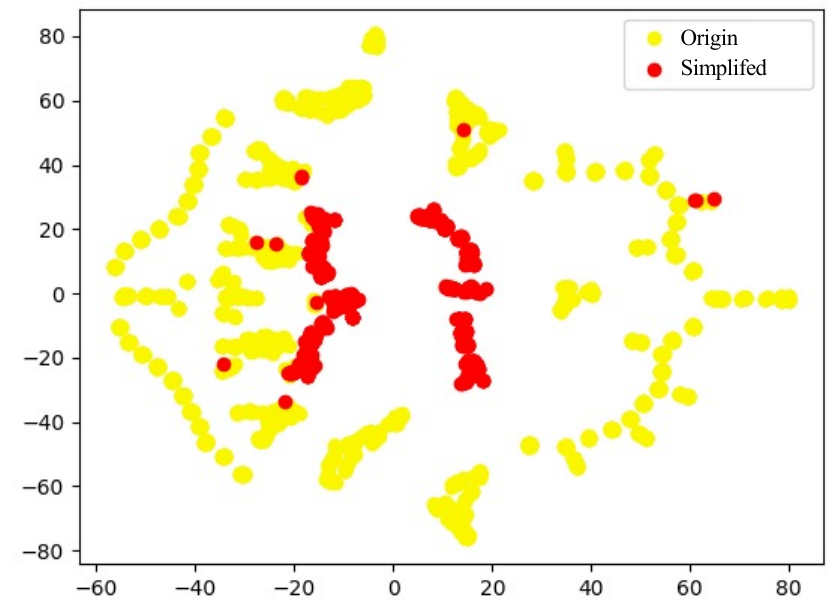}
        \caption{Coordinates Distribution}
    \end{subfigure}
    \begin{subfigure}[b]{0.49\linewidth}
        \includegraphics[width=\linewidth]{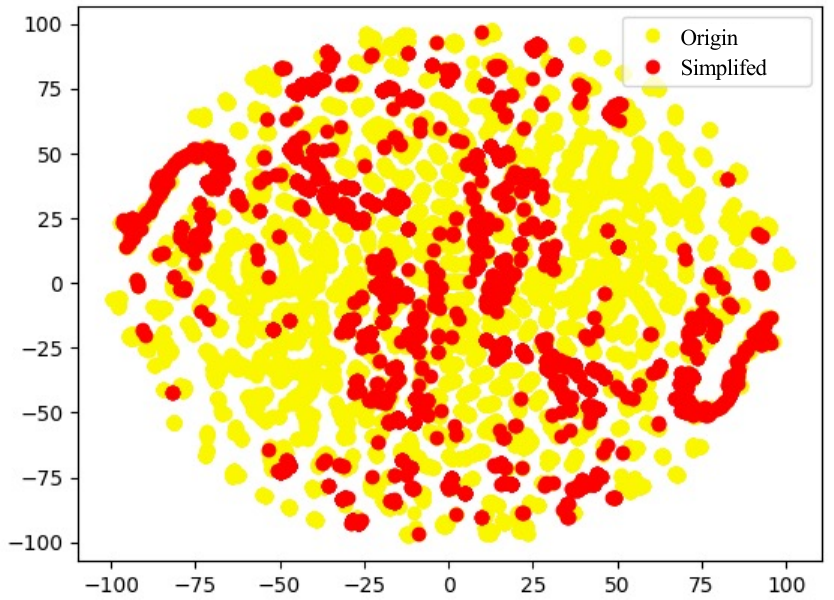}
        \caption{Relation Distribution}
    \end{subfigure}
    \caption{t-SNE \cite{van2008tsne} visualization of the distribution between original and simplified meshes. (a) shows the existence of distribution shifts between the original and decimated coordinates. However, (b) shows that the distributions of decimated relations $D(v_i, v_j)$ are included in the distributions of the original ones.}
    \label{fig:method_distribution_relation_xyz}
\end{figure}

Graph attention network (GAT) is a convolution operator defined on graphs. For $l$-th layer of GAT, the input is a set of vertex features $\mathbf{F}^l=\{F^l_{v_0}, F^l_{v_2}, ..., F^l_{v_N}\}$, where $N$ is the number of vertices. This layer produces a new set of vertex features $\mathbf{F}^{l+1}=\{F^{l+1}_{v_0}, F^{l+1}_{v_2}, ..., F^{l+1}_{v_N}\}$. For each vertex $v_i$, its new feature $F^{l+1}_{v_i}$ is computed as the averaged weighted sum of its neighbor features $F^{l}_{v_j}$ for all $v_j\in \mathcal{N}(v_i)$. To increase expressive power, weights for each neighbor $v_j$ are obtained from learnable linear transform $W_{\Theta}$.

We view 3D meshes as graphs $G(V,E)$. We first define our GAT on meshes as:
\begin{equation}
    \begin{gathered}
        F^{l+1}_{v_i}=\frac{1}{|\mathcal{N}(v_i)|}\sum_{v_j\in \mathcal{N}(v_i)}F^l_{v_j}W_{\Theta}(D(v_j,v_i))
    \end{gathered}
    \label{eq:3Dconv}
\end{equation}
where $F^{l+1}(v_i)$ is the feature vector of $v_i$ at $(l+1)$-th layer. The neighborhood $\mathcal{N}(v_i)=\{v_j| (v_j, v_i)\in E\}\cup \{v_i\}$ is defined as all points adjacent to the point $v_i$ and itself. We use a multilayer perceptron (MLP) to model the learnable linear transform $W_{\Theta}$. The input of the MLP is the relation $D(v_i, v_j)$ between the target vertex $v_i$ and its neighbor $v_j$. Figure \ref{fig:GAT} gives a visualization process of our GAT.

We show that it is beneficial to learn from relation $D(v_i, v_j)$. We define relation $D(v_i, v_j)=\Vec{v_i}-\Vec{v_j}$ as the coordinate offset from $v_i$ to $v_j$. Such the relation can survive the mesh simplification algorithm \cite{lindstrom1998decimation1, lindstrom1999decimation2}. First, we simplify meshes to reduce the vertex number to $1/5$ of the original, \ie $N_v'=1/5 N_v$. Then we visualize the coordinates and relation distributions for both original and simplified meshes in Figure \ref{fig:method_distribution_relation_xyz}. Figure \ref{fig:method_distribution_relation_xyz} shows coordinate distribution differences between original and simplified meshes, while relation distributions are still included in the original distributions. Based on this insight, our method is trained on simplified meshes and shows adaptability to increased-size meshes.

\subsection{Deep3DMark} \label{sec:arch}
Our architecture (Figure \ref{fig:arch}) consists of five parameterized learnable components: 1) a message autoencoder that can map a binary message $M$ to a latent code $z$ and decodes $z$ back to $M$, 2) an encoder $\mathbf{E}$ models function $\mathcal{E}_{\phi}$ that generates a watermarked vertices $\hat{V}$ given $V$ and $z$, 3) an attack layer applies perturbation over $\hat{V}$ to increase the robustness in the way of data augmentation, 4) a decoder $\mathbf{D}$ models $\mathcal{D}_{\theta}$ that reconstructs binary message from $\hat{v}$, and 5) a discriminator $\mathbf{A}$ encourages $\hat{V}$ indistinguishable from $V$.

The \textbf{encoder} $\mathbf{E}$ first applies convolutions to input $V$ to form some intermediate representation. We aim to incorporate message latent code $z$ in the way that the encoder learns to embed parts of it at any spatial location of $V$. To achieve this, we replicate the latent code and concatenate it to the intermediate representations. We apply more convolutions to transform the concatenated feature to watermarked vertices $\hat{V}$. 

The \textbf{attack layer} applies perturbations to generated $\hat{V}$. The perturbations consider several mesh attacks, including 1) Gaussian noise with mean $\mu$ and deviation $\sigma$, 2) random rotation attacks with rotate center $(x,y,z)$ and degree $\alpha$, 3) translation attack and 4) scaling attack with a scaling ratio $s$. Our ablation study shows that the attack layer effectively increases the robustness against multiple attacks.

The \textbf{decoder} $\mathbf{D}$ first applies several convolutions to generate the intermediate representation of $\hat{V}$. It finally uses a global average pooling followed by an MLP layer to generate a vector of the same size as the latent code $z$. The global average pooling layer ensures that our method aggregates information from all vertices.

The \textbf{adversarial discriminator} $\mathbf{A}$ shares a similar structure as the decoder except that its final MLP layer transforms the aggregated vector into a binary classification, which indicates whether the given $\hat{V}$ is generated by the encoder $\mathbf{E}$.

According to Shannon's capacity theory \cite{shannon1948mathematical}, redundancy is necessary to achieve robustness. The \textbf{message autoencoder} increases the robustness of our system by injecting redundancy into the system. Given a binary message $M$ of length $N_m$, the message encoder maps it into a latent code $z$ of length $N_z>N_m$, which can be used to recover $M$ through a message decoder. We train the autoencoder in a way that the decoder can recover $M$ from the noised latent code $\hat{z}$. We choose NECST \cite{choi2019necst}, a learnable channel coding method, as our message autoencoder. Our message autoencoder is trained independently from the entire watermarking model.

\subsection{Training and Losses} \label{sec: loss}

\begin{table*}
    \centering
    \small
    \begin{tabular}{c|c|c|c|c|c|c|c|c|c|c|c}
    \toprule[1pt]
    \multirow{2}{*}{Algorithm} & \multirow{2}{*}{Category} & \multicolumn{3}{c|}{Geometry Difference} & \multicolumn{6}{c|}{Accuracy \& Robustness (\%)} & \multirow{2}{*}{Time (s)}\\\cline{3-11}
    &  & $L_1d$ & Hausdorff & SNR & w/o attack & Gauss & Trans & Rot & Scale & Crop &\\ \hline\hline
    \jiang & \multirow{7}{*}{Traditional} & 0.1051 & 1.3950 & 39.50 & 80.59 & 76.98 & 48.2 & 80.58 & 49.97 & 49.68 & 81.24\\
    \spectral & & 0.0090 & 0.0125 & 44.99 & 87.48 & 54.31 & 50.41 & 50.40 & 54.29 & 56.01 & 70.350\\
    \peng & & 0.0182 & 0.0158 & 46.70 & 97.98 & 50.24 & 58.30 & 50.55 & 50.25 & 50.30 & 30.470\\
    \ppeng & & 0.0073 & 0.0375 & 36.94 & 80.89 & 79.28& 80.89 & 80.89 & 80.89 & 54.12 & 3.240\\
    \tsai & & 0.0101 & 0.0145 & 40.90 & 96.41 & 50.03 & 49.99 & 49.75 & 50.02 & 50.10 & 5.989\\
    \ttsai & & \textbf{0} & \textbf{0} & \textbf{inf} & 84.49 & 84.48 & 84.48 & 64.32 & 61.76 & 50.51 & 3.079\\
    \hou & & 0.0194 & 0.0268 & 45.10 & 90.84 & 90.84 & 69.22 & 87.92 & 49.55 & 50.21 & 2.090\\

    \hline
    \wang & \multirow{2}{*}{DNN-based} & 0.1337 & 0.1856 & 15.04 & 97.50 & 90.84 & 90.22 & 90.92 & 90.55 & 70.21 & 0.010\\
    \tool \textbf{(Ours}) & & 0.0338 & 0.0617 & 30.84 & \textbf{98.17} & \textbf{91.06} & \textbf{98.17} & \textbf{96.84} & \textbf{98.17} & \textbf{78.90} & \textbf{0.0089}\\
     
    \bottomrule[1pt]
    \end{tabular}
    \caption{Comparison of watermarked mesh quality (geometry difference), reconstruction accuracy without attack (w/o attack), robustness under Gaussian noise (Gauss), translation (Trans), rotation (Rot), scaling (Scale), and cropping attack (Crop).}
    \label{tab:compare_previous}
\end{table*}

We achieve the objective in Eq \ref{eq:finalobj} using three losses: encoding loss $L_{enc}$, reconstruction loss $L_{dec}$, and discriminative loss $L_{dis}$. Formally:
\begin{equation}
    \begin{aligned}
        &\phi^*, \theta^*=\\
        &\argmin_{\phi, \theta}(\lambda_{enc} L_{enc}(\phi, \theta) + \lambda_{dec} L_{dec}(\phi, \theta) + \lambda_{dis} L_{dis}(\phi, \theta))
    \end{aligned}
\end{equation}
where $\lambda_{enc},\lambda_{dec},\lambda_{dis}$ are weight factors. Both $L_{enc}, L_{dis}$ encourage generated $\hat{V}$ indistinguishable from $V$. For $L_{enc}$, we use both the L2 norm and infinite norm of geometry difference to penalize the distortion:
\begin{equation}
    L_{enc}=\frac{1}{N_v}\sum_{i}^{N_v}(V[i] - \hat{V}[i])^2 + \max_i\{V[i]-\hat{V}[i]\}
\end{equation}
For $L_{dis}$, we use part of sigmoid cross entropy loss:
\begin{equation}
    L_{dis}=\log(1-\sigma(\mathbf{A}(\hat{V})))
\end{equation}
We apply standard sigmoid cross entropy loss to encourage precise message reconstruction:
\begin{equation}
    \begin{aligned}
        &L_{dec}=\\
        &\frac{\sum_{i}^{N_m}(M[i]\cdot\log\sigma(\hat{M}[i]) + (1-M[i])\cdot\log(1-\sigma(\hat{M}[i])))}{N_m}
    \end{aligned}
\end{equation}

The final message bits are computed from the following:
\begin{equation}
    M_{final}=clamp(sign(\hat{M}-0.5), 0, 1)
\end{equation}

%% file: chapter/experiment.tex
\section{Experiment}

\subsection{Experiment Setup}
\textbf{Training Settings.} Our experiment is conducted on Ubuntu 18.04, with 503GB RAM and five Nvidia RTX 3090. Our \tool uses GATs to build $\mathbf{E}$, $\mathbf{D}$ and $\mathbf{A}$, where channel size are all 64. At the first layer, we take coordinates $(x,y,z)$ as the feature of points, i.e., $C_v=3$. Our experiment for both \tool and other baselines are evaluated under the message length $N_m=8$. During training, we set $\lambda_{enc}=2,\lambda_{dec}=1,\lambda_{dis}=0.001$ under the settings of 8-bit message lengths, and we set $\mu=0$, $\sigma=0.001$, $\alpha\in[0,\pi)$, $s\in[0.1,1)$.

\noindent\textbf{Baselines.} We adopt nine state-of-the-art watermarking algorithms as our baselines. \jiang \cite{jiang2017reversible}, \ppeng \cite{peng2022semi}, \tsai \cite{tsai2020separable}, \ttsai \cite{tsai2022integrating} and \hou \cite{hou2023separable} are encrypted domain watermarking algorithms, whose geometry difference can only be evaluated after model decryption. We evaluate the geometry difference between the original and decrypted watermarked mesh. \peng \cite{peng2021general}, \spectral \cite{al2019graph} and \wang \cite{wang2022deep} are plain-text domain watermarking algorithms whose geometry difference can be directly evaluated between the original and watermarked mesh. We also compare with \yoo \cite{yoo2022deep} in geometry difference and accuracy.

\noindent\textbf{Metrics.} To evaluate the extracted message accuracy, we use average bit accuracy. To evaluate geometry differences, we use Hausdorff distance, the L1 norm of vertex difference ($L_1d$), and signal-to-noise ratio (SNR).

\subsection{Dataset}
Our \tool is trained on a simplified train set from ModelNet40 \cite{modelnet} and then tested on the entire test of ModelNet40 and other datasets such as ShapeNet \cite{shapenet2015}, GraspNet \cite{fang2020graspnet}, ScanNet \cite{dai2017scannet} and Hands \cite{romero2022hands}. For all datasets, we normalize the vertex coordinates $(x,y,z)$ to $[-1,1]$ before meshes are fed into the network unless explicitly mentioned.

We acquire simplified data through a simplification using CGAL \cite{cgal:eb-22b}, which performs edge-collapse or half-edge-collapse algorithms to reduce the number of triangles by merging vertices. We generated two train sets \textit{m500} and \textit{m2500}. The number of vertices in \textit{m500} and \textit{m2500} are $N_v=500$ and $N_v=2500$, respectively. For \textit{m2500}, we manually filter out meshes whose $N_v$ is originally less than 2500 and those with low quality after simplifications. We also perform the same process for \textit{m500}. As a result, we get 3508 train meshes and 879 test meshes for \textit{m2500}, and 1147 train meshes and 337 test meshes for \textit{m500}. The original ModelNet has 9843 and 2468 meshes for training and testing. We train two replicas of \tool on \textit{m500} and \textit{m2500}, respectively. Both are further tested on the test set of ModelNet to evaluate the size adaptability.

To evaluate the effectiveness on geometry variations, two replicas of \tool, which is trained on \textit{m500} and \textit{m2500}, are tested on ShapeNet, GraspNet, ScanNet, and Hands. ShapeNet has different categories of meshes from ModelNet, such as birdhouse, camera, clock, etc. Scannet is a dataset of scanned and reconstructed real-world scenes. Hands contain meshes of human hands.

\begin{figure*}[htb]
    \centering
    \begin{subfigure}[b]{0.21\linewidth}
        \includegraphics[width=\linewidth, page=1]{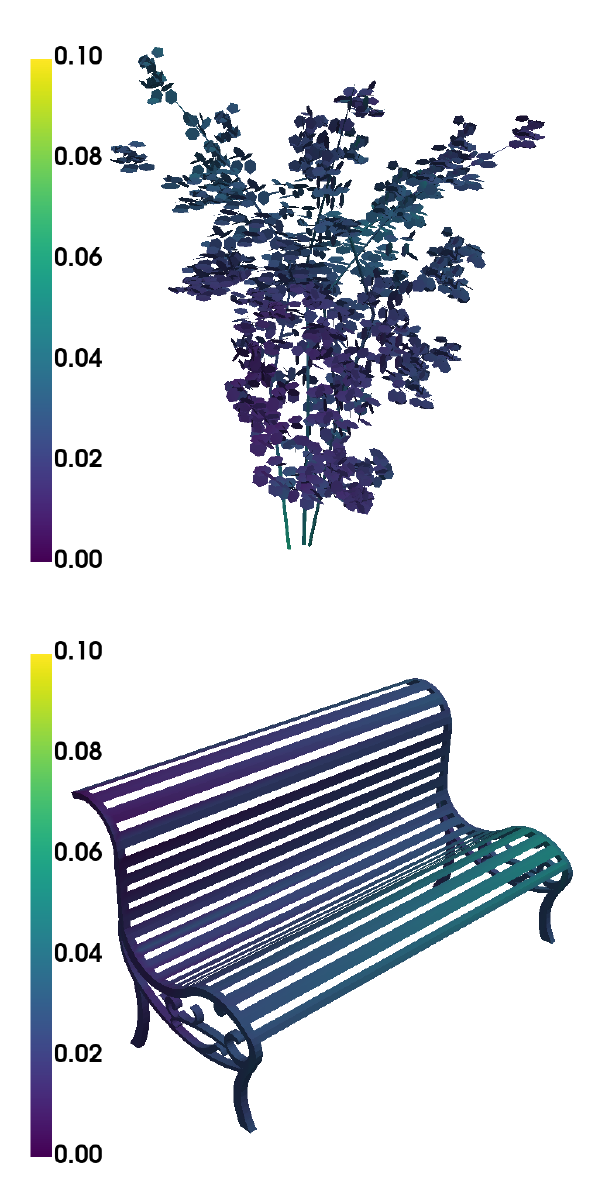}
        \caption{Visualization $L_1d$}
    \end{subfigure}
    \begin{subfigure}[b]{0.21\linewidth}
        \includegraphics[width=\linewidth, page=2]{figures/experiments/qualitative/method_qualitative.pdf}
        \caption{Original Mesh}
    \end{subfigure}
    \begin{subfigure}[b]{0.21\linewidth}
        \includegraphics[width=\linewidth, page=3]{figures/experiments/qualitative/method_qualitative.pdf}
        \caption{Watermarked Mesh}
    \end{subfigure}
    
    \caption{Results to show the imperceptible watermarking: (a) L1 Norm of vertex difference ($L_1d$), (b) the original mesh, and (c) the watermarked mesh with \tool (ours).}
    \label{fig: mesh_quality}
\end{figure*}

\subsection{Experiment on Robustness} \label{sec:robust}

\noindent \textbf{Settings.} Our first experiment is to (1) prove that it is hard to achieve full robustness even under the settings of 8-bit message length for traditional methods and (2) \tool is robust against Gaussian, rotation, scaling, translation, and cropping attack while maintaining relatively high quality. \tool is trained on the train set of \textit{m2500}. \tool and baselines are further evaluated on the test set of \textit{m2500}. For a fair comparison, the watermarked meshes are rescaled back to their original coordinates before we apply any attacks. We choose Gaussian noise ($\sigma=0.1$), translation (random translation vector in $[0,1000]^3$), rotation (origin point as the rotation center and $\alpha\in [0,\frac{\pi}{2}]$, ), scaling ($s\in[0.1,1)$) and cropping attacks (cropping ratio $c=0.1$).

\noindent \textbf{Results.} Table \ref{tab:compare_previous} shows that although traditional methods have relatively high watermarked mesh quality, they are vulnerable to multiple attacks. \peng embeds secret messages in the least significant bits of vertex coordinates, thus vulnerable to Gaussian attacks. \ttsai embeds secret messages in the most significant bits of vertex coordinates. However, it still cannot withstand rotation and scaling attacks. Compared to traditional methods, \tool is robust against arbitrary attacks and achieved 10\%$\sim$50\% higher accuracy when facing arbitrary attacks. Compared to DNN-based methods, \tool achieved 1\%$\sim$8\% higher accuracy and 2$\times$ SNR value. Table \ref{tab:compare_yoo} shows that we achieve similar geometry difference and accuracy to \yoo. We provide a visual quality example in Fig. \ref{fig: mesh_quality} for the concern of the SNR drop compared with traditional methods.

\begin{table}[htb]
    \centering
    \small
    \begin{tabular}{c|c|c}
    \toprule[1pt]
    Algorithm & L1 Vertex Normal & Bit Accuracy (\%)\\
    \hline
    \yoo & \textbf{0.1041} & 0.9362 \\
    \tool & 0.1143 & \textbf{0.9403} \\
    
    \bottomrule[1pt]
    \end{tabular}
    \caption{Comparison with \yoo.}
    \label{tab:compare_yoo}
\end{table}

\begin{table}[htb]
    \centering
    \small
    \begin{tabular}{c|c}
    \toprule[1pt]
    Distortions Made By & Bit Accuracy (\%)\\
    \hline
    \hline
    No Distortion & 98.17 \\
    \hline
    Gaussian Noise ($\sigma=0.005$) & 98.06 \\
    Gaussian Noise ($\sigma=0.01$) & 97.11 \\
    Gaussian Noise ($\sigma=0.02$) & 90.47 \\
    \hline
    Cropping ($c=0.1$) & 78.41\\
    Cropping ($c=0.5$)& 75.04\\
    Cropping ($c=0.9$)& 65.81\\
    \hline
    Implicit Laplacian Smooth ($\lambda_I=1.0$) & 93.57 \\
    Implicit Laplacian Smooth ($\lambda_I=5.0$) & 89.25 \\
    Implicit Laplacian Smooth ($\lambda_I=10$) & 80.29 \\
    \hline
    Draco Compression($N_q=15$) & 97.32\\
    Draco Compression($N_q=10$) & 96.98\\
    Draco Compression($N_q=5$) & 88.60\\
    \hline
    Reorder & 98.17\\
    ARAP & 96.42\\
    \bottomrule[1pt]
    \end{tabular}
    \caption{The robustness under unknown attacks, where Gaussian noise is tested with out-of-domain parameters.}
    \label{tab:attack}
\end{table}

\subsection{Experiment on Unknown Distortions}
\noindent \textbf{Settings.} Our second experiment is to prove \tool is robust against unknown distortions because a practical watermarking experiment must be robust against a wide range of distortions. We choose Gaussian noise, cropping, reordering, ARAP \cite{sorkine2007rigid}, implicit laplacian smoothing (Smooth) \cite{desbrun2023implicit} and Draco \cite{draco1} compression. For the ARAP attack, we randomly select $[0,10]$ handle points and move all the handle points along a vector with length $0.1$. For implicit laplacian smoothing, we increase the distortion strength by increasing the parameter $\lambda_I$. For Draco compression, we increase the distortion strength by decreasing the quantization bits $N_q$.

\noindent \textbf{Results.} Table \ref{tab:attack} shows the bit accuracy of our model on these additional distortions. Overall, our model shows full robustness on these unknown distortions.

\begin{table*}[htb]
    \centering
    \small
    \begin{tabular}{c|c|c|ccccccccc}
    \toprule[1pt]
     & Metrics & Avg & birdhouse & camera & clock & spigot & knife & loudspeaker & mug & pistol & printer \\\hline \hline
     \multirow{3}{*}{\textit{m500}} & Hausdorff & 0.0758 & 0.0778 & 0.0723 & 0.0754 & 0.0762 & 0.0747 & 0.0690 & 0.0882 & 0.0809 & 0.0794\\
      & $L_1d$ & 0.0308 & 0.0349 & 0.0339 & 0.0476 & 0.0364 & 0.0355 & 0.0421 & 0.0313 & 0.0380 & 0.0333\\
      & SNR & 26.54 & 27.91 & 27.99 & 26.08 & 25.79 & 25.22 & 26.58 & 26.66 & 27.67 & 26.18\\
      & Acc & 0.8863 & 0.8698 & 0.9347 & 0.9220 & 0.7943 & 0.7160 & 0.9221 & 0.8615 & 0.8969 & 0.9390\\
    \hline
     \multirow{3}{*}{\textit{m2500}} & Hausdorff & 0.0549 & 0.0680 & 0.0568 & 0.0570 & 0.0474 & 0.0493 & 0.0620 & 0.0527 & 0.0562 & 0.0547 \\
      & $L_1d$ & 0.0248 & 0.0245 & 0.0302 & 0.0278 & 0.0281 & 0.0267 & 0.0293 & 0.0266 & 0.0254 & 0.0296\\
      & SNR & 31.01 & 29.92 & 27.55 & 31.11 & 27.14 & 30.30 & 27.02 & 30.68 & 30.76 & 31.19 \\
      & Acc & 0.9348 & 0.9554 & 0.9800 & 0.9489 & 0.9358 & 0.8773 & 0.9682 & 0.9526 & 0.9324 & 0.9623 \\
    \bottomrule[1pt]
    \end{tabular}
    \caption{Geometry adaptability on ShapeNet dataset. \textit{m500} and \textit{m2500} are trained on simplified ModelNet dataset with $N_v=500$ and $N_v=2500$, respectively. Here, we list the results of nine categories.}
    \label{tab:adaption_shapenet}
\end{table*}

\begin{table*}
    \centering
    \small
    \begin{tabular}{c|c|c|c|c|c|c}
    \toprule[1pt]
     & Metrics & $(0,20000)$ & $[20000,40000)$ & $[40000,60000)$ & $[60000,80000)$ & $[80000,100000)$\\
    \hline\hline
    \multirow{3}{*}{\textit{m500}} & Hausdorff & 0.0721 & 0.0609 & 0.0600 & 0.0626 & 0.0666\\
      & $L_1d$ & 0.0408 & 0.0344 & 0.0348 & 0.0355 & 0.0304\\
      & SNR & 26.65 & 27.43 & 26.61 & 26.552 & 26.91\\
      & Acc & 0.9183 & 0.9052 & 0.8570 & 0.8717 & 0.8181\\\hline

    \multirow{3}{*}{\textit{m2500}} & Hausdorff & 0.0550 & 0.0473 & 0.0466 & 0.0502 & 0.0488\\
      & $L_1d$ & 0.0281 & 0.0225 & 0.0232 & 0.0232 & 0.0222\\
      & SNR & 29.83 & 31.21 & 30.18 & 30.33 & 30.02\\
      & Acc & 0.9462 & 0.9034 & 0.9183 & 0.9123 & 0.8708\\
    \bottomrule[1pt]
    \end{tabular}
    \caption{Size adaptability on ModelNet dataset with the varied number of vertices $N_v\in(0,100000]$. \textit{m500} and \textit{m2500} are trained on simplified ModelNet dataset with $N_v=500$ and $N_v=2500$, respectively.}
    \label{tab:adaption_size}
\end{table*}

\subsection{Experiment on Adaptability} \label{sec:transfer}

\begin{table}[htb]
    \centering
    \small
    \begin{tabular}{c|c|ccc}
    \toprule[1pt]
     & Metrics & GraspNet & Hands & ScanNet\\\hline \hline
     \multirow{3}{*}{\textit{m500}} & Hausdorff & 0.0817 & 0.0753 & 0.0970\\
      & $L_1d$ & 0.0365 & 0.0337 & 0.0378\\
      & SNR & 28.80 & 28.91 & 27.07\\
      & Acc & 0.9588 & 0.9288 & 0.8593\\
    \hline
    \multirow{3}{*}{\textit{m2500}} & Hausdorff & 0.0519 & 0.0515 & 0.0527\\
      & $L_1d$ & 0.0254 & 0.0277 & 0.0237\\
      & SNR & 30.62 & 30.73 & 30.93\\
      & Acc & 0.9673 & 0.9584 & 0.9929\\
    \bottomrule[1pt]
    \end{tabular}
    \caption{Geometry adaptability on other datasets.}
    \label{tab:adaption_others}
\end{table}

\begin{figure}[htb]
    \centering
    \begin{subfigure}[b]{0.19\linewidth}
        \includegraphics[width=\linewidth]{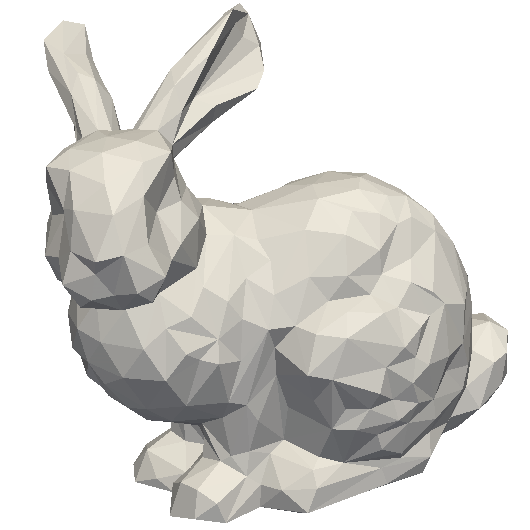}
        \caption{500}
    \end{subfigure}
    \begin{subfigure}[b]{0.19\linewidth}
        \includegraphics[width=\linewidth]{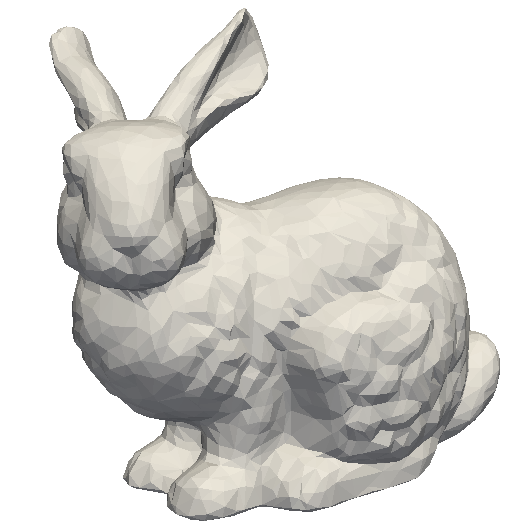}
        \caption{1000}
    \end{subfigure}
    \begin{subfigure}[b]{0.19\linewidth}
        \includegraphics[width=\linewidth]{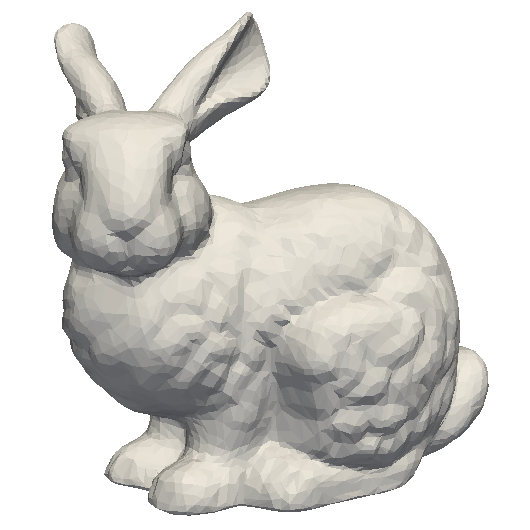}
        \caption{2500}
    \end{subfigure}
    \begin{subfigure}[b]{0.19\linewidth}
        \includegraphics[width=\linewidth]{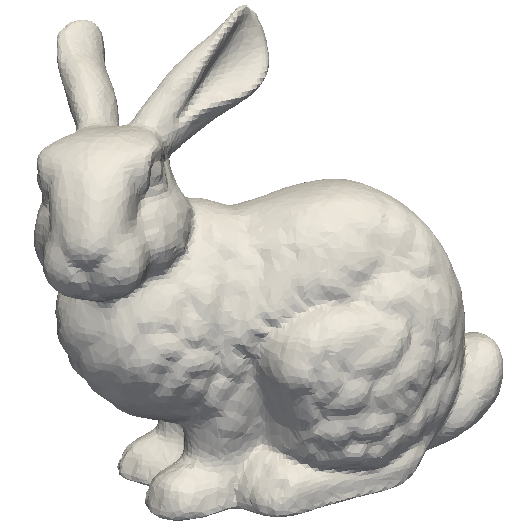}
        \caption{10000}
    \end{subfigure}
    \begin{subfigure}[b]{0.19\linewidth}
        \includegraphics[width=\linewidth]{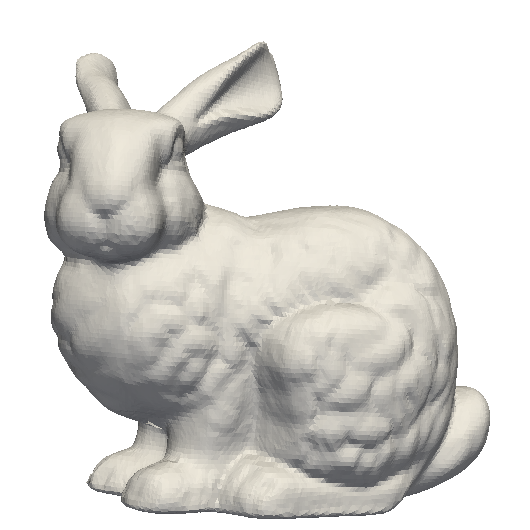}
        \caption{35947}
    \end{subfigure}
    \\
    \begin{subfigure}[b]{0.19\linewidth}
        \includegraphics[width=\linewidth]{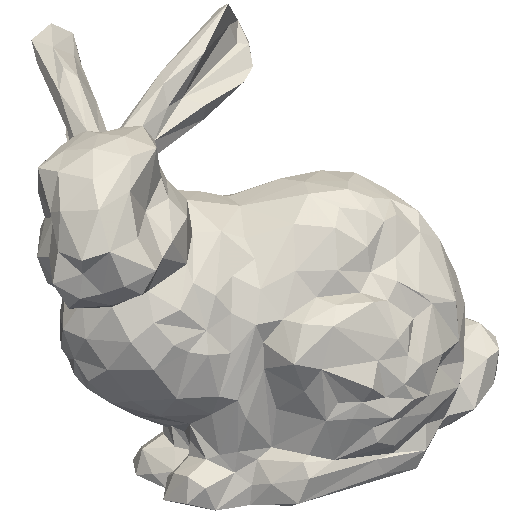}
        \caption{500}
    \end{subfigure}
    \begin{subfigure}[b]{0.19\linewidth}
        \includegraphics[width=\linewidth]{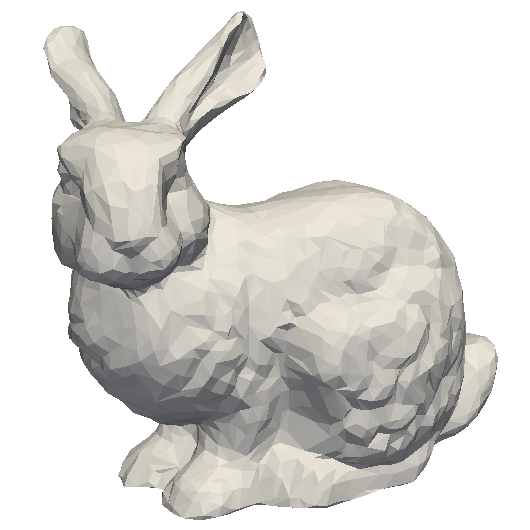}
        \caption{1000}
    \end{subfigure}
    \begin{subfigure}[b]{0.19\linewidth}
        \includegraphics[width=\linewidth]{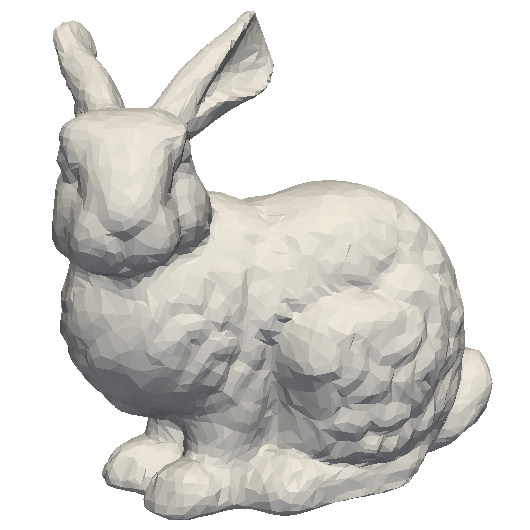}
        \caption{2500}
    \end{subfigure}
    \begin{subfigure}[b]{0.19\linewidth}
        \includegraphics[width=\linewidth]{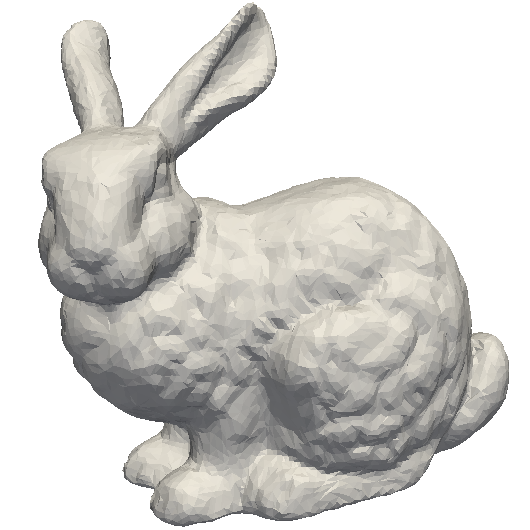}
        \caption{10000}
    \end{subfigure}
    \begin{subfigure}[b]{0.19\linewidth}
        \includegraphics[width=\linewidth]{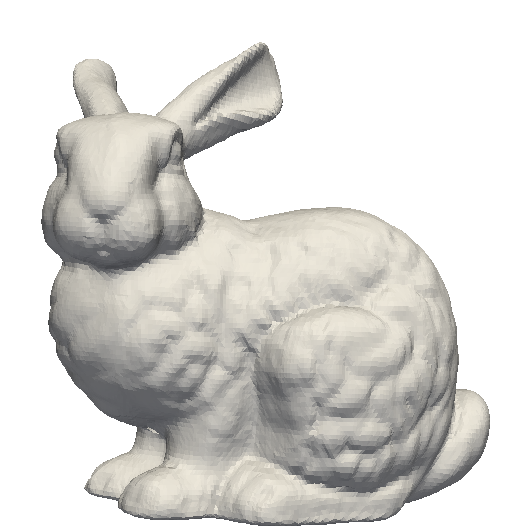}
        \caption{35947}
    \end{subfigure}
    \caption{Results of the \tool trained on \textit{m2500} under different mesh sizes. \textbf{Top(a-e)}: original mesh $G$ with varying $N_v$ from 500 to 35947. \textbf{Bottom(f-j)}: the corresponding watermarked mesh $\hat{G}$ using our \tool. (a-d) are the simplified version of the original mesh (e).}
    \label{fig:qualitative_wm_gan_2500}
\end{figure}

\noindent \textbf{Settings.} Our third experiment is conducted to prove that our method can be generalized to different mesh sizes and unseen geometry. We train \tool on \textit{m500} and \textit{m2500} to get two replicas. Both are further evaluated on the original test set of ModelNet40, ShapeNet40, ScanNet, GraspNet, and Hands. The data distributions in ShapeNet40, ScanNet, GraspNet, and Hands are unseen to both \tool replicas during training.

\noindent \textbf{Results.} Figure \ref{fig:qualitative_wm_gan_2500} shows the result using the \tool trained on \textit{m2500}. The top row shows the cover meshes where (a-d) are simplified from the original mesh (e). The bottom row shows the watermarked meshes. Table \ref{tab:adaption_size} shows statistical results under size variations. We evaluate our method on meshes with $N_v\leq 100000$. The \tool trained on \textit{m2500} is still effective when mesh size is 40$\times$ increased. Compared to the \tool trained on \textit{m2500}, the one trained on \textit{m500} achieves lower accuracy and introduces more distortions. However, it still achieves 81.81\% accuracy when the mesh size is 190$\times$ increased.
Table \ref{tab:adaption_shapenet} shows results under geometry variations on ShapeNet. On ShapeNet, the \tool trained on \textit{m2500} achieves an average 93.48\% bit accuracy while only introducing 0.0248 L1 norm of vertex difference and 0.0549 Hausdorff difference. The results demonstrate that simplified meshes inherit the relation $D(v_i, v_j)$ distribution from the original meshes. However, as the size of training meshes decreases, the adaptability of GAT decreases as well.

\section{Conclusion}
3D watermarking is a key step toward copyright protection. Our paper has introduced \tool, which utilizes graph attention networks to embed binary messages in vertex distributions without texture assistance. Our approach has taken advantage of the property that simplified meshes inherit similar relations from the original ones, specifically the offset vector between adjacent vertices. This approach has enabled the training on simplified meshes but remains effective on larger and previously unseen categories of meshes (adaptability), resulting in fewer distortions and 10\%$\sim$50\% higher bit accuracy than previous methods when facing attacks. Moreover, extensive experiments have shown that our \tool is robust against unknown mesh attacks, such as smoothing, ARAP, and compression. 


%% file: chapter/supplementary.tex
\section{Architecture}

\begin{figure}[htb]
    \centering
    \includegraphics[width=\linewidth, page=3]{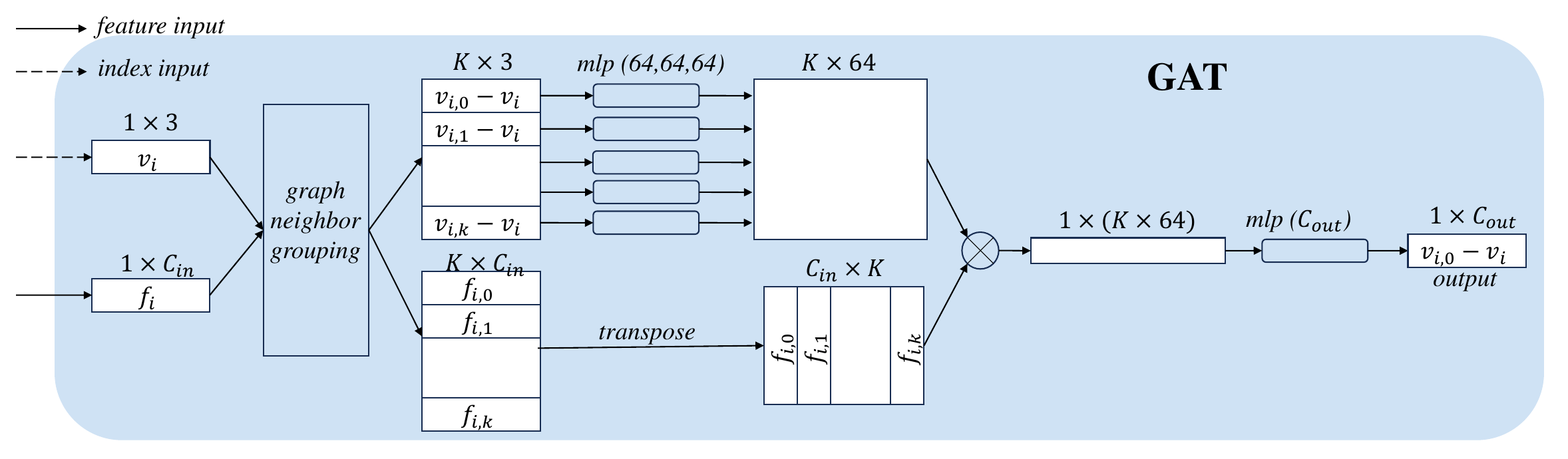}
    \caption{GAT Details}
    \label{fig:GAT}
\end{figure}

\begin{figure}[htb]
    \centering
    \includegraphics[width=\linewidth, page=1]{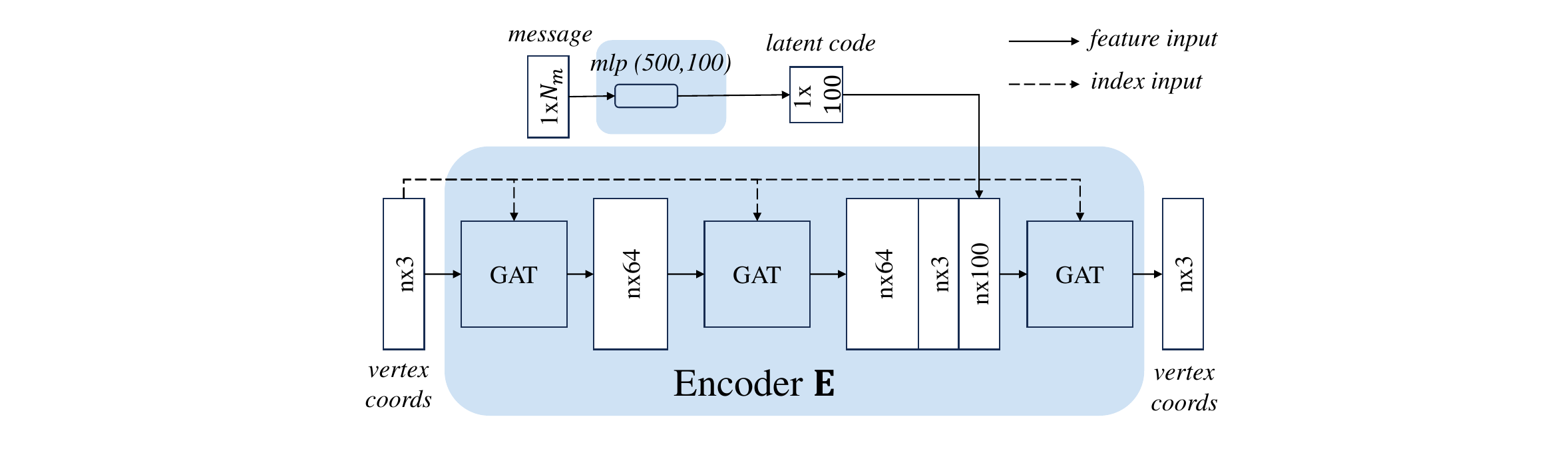}
    \caption{Encoder Details}
    \label{fig:encoder}
\end{figure}

\begin{figure}[htb]
    \centering
    \includegraphics[width=\linewidth, page=2]{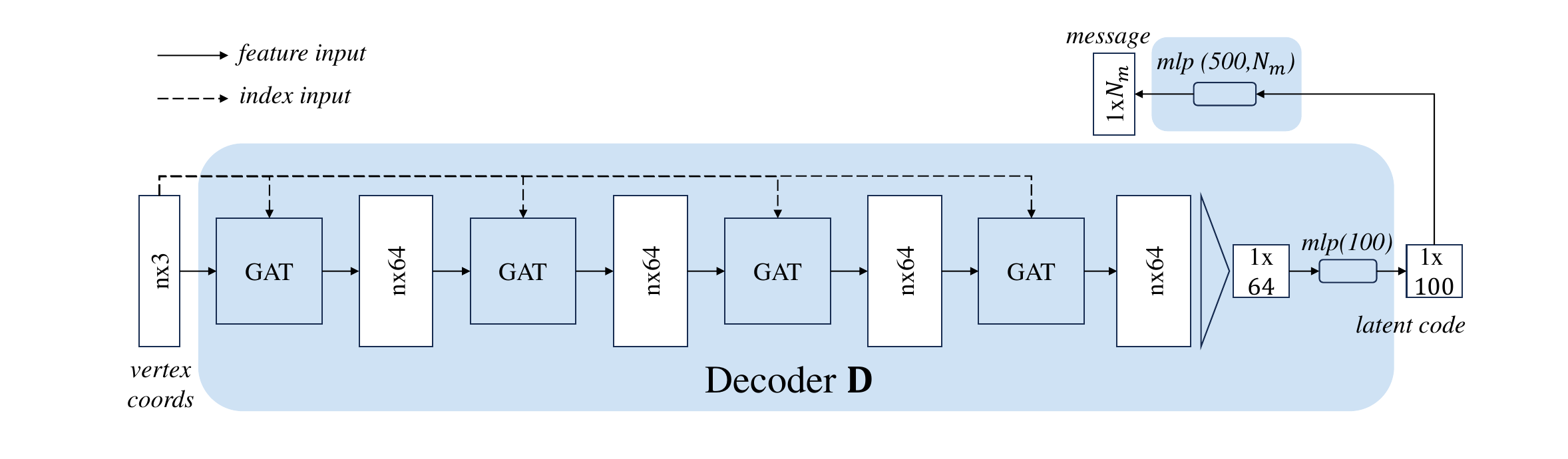}
    \caption{Decoder Details}
    \label{fig:decoder}
\end{figure}

Figure \ref{fig:GAT} shows our GAT architecture (GAT), which is used as our backbone. For efficient computing, we use the reformulation proposed in PointConv \cite{wu2019pointconv}, significantly improving the GAT performance. We use three layers of GATs to build encoder $\mathbf{E}$ (Fig. \ref{fig:encoder}) and four layers of GATs to build decoder $\mathbf{D}$ (Fig. \ref{fig:decoder}) and discriminator $\mathbf{A}$. The only difference between $\mathbf{D}$ and $\mathbf{A}$ is that the last MLP layer of $\mathbf{D}$ transforms the aggregated vector into latent code of decoded secret message while $\mathbf{A}$ transforms the aggregated vector into binary classifications.


\section{Complexity}
\begin{figure}[htb]
    \centering
    \includegraphics[width=\linewidth]{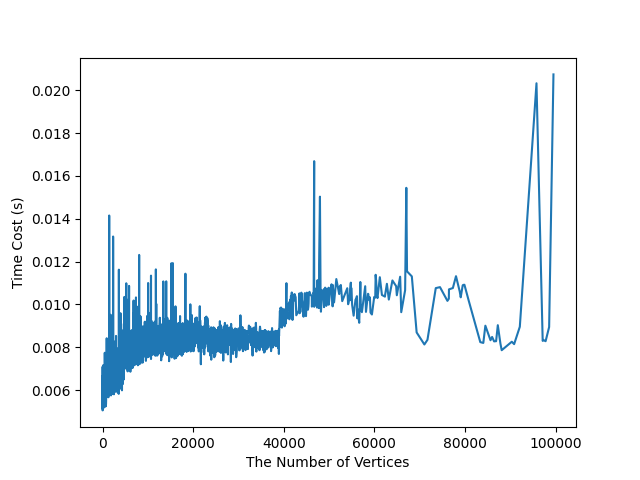}
    \caption{Time growth with the increase of $N_v$}
    \label{fig:time_growth}
\end{figure}

We show our time growth with increasing vertices in Fig. \ref{fig:time_growth}. Our model consists of only 2.92 million parameters.

\section{More Visual Results}
\begin{figure*}[htb]
    \centering
    \begin{subfigure}[b]{0.33\linewidth}
        \includegraphics[width=\linewidth, page=1]{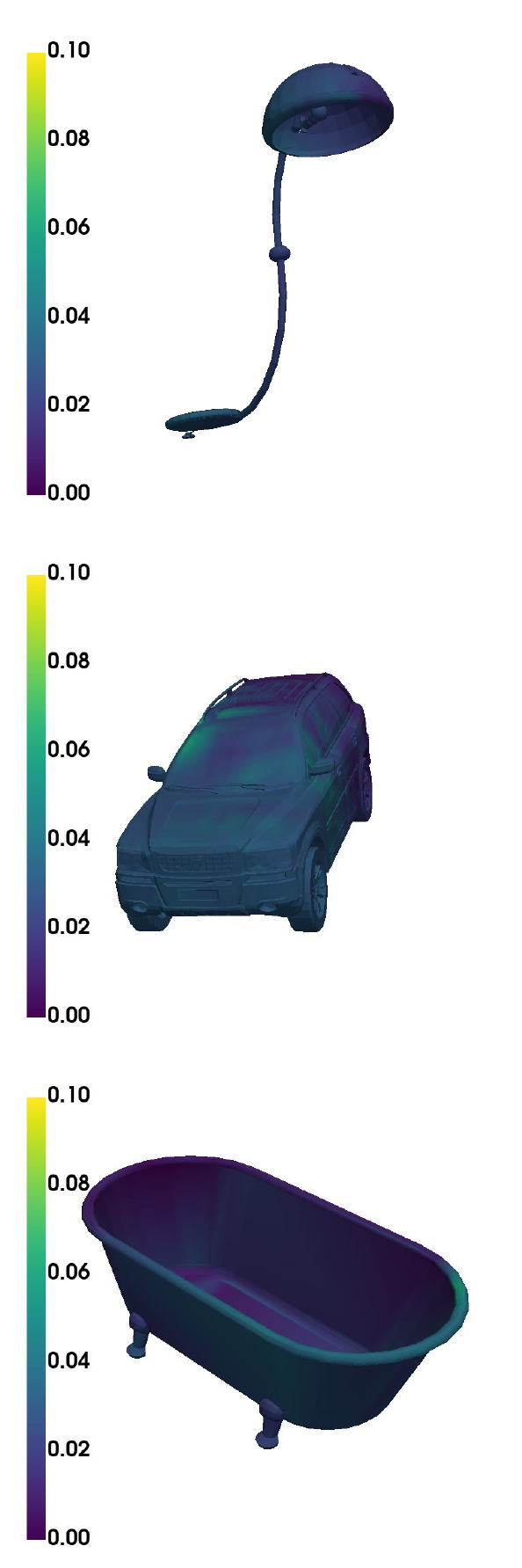}
        \caption{Visualization $L_1d$}
    \end{subfigure}
    \begin{subfigure}[b]{0.33\linewidth}
        \includegraphics[width=\linewidth, page=2]{figures/experiments/qualitative/more_visual.pdf}
        \caption{Original Mesh}
    \end{subfigure}
    \begin{subfigure}[b]{0.33\linewidth}
        \includegraphics[width=\linewidth, page=3]{figures/experiments/qualitative/more_visual.pdf}
        \caption{Watermarked Mesh}
    \end{subfigure}
    
    \caption{ModelNet40 Results to show the imperceptible watermarking: (a) L1 Norm of vertex difference ($L_1d$), (b) the original mesh, and (c) the watermarked mesh with \tool (ours). The visual data are lamp\_0052, car\_0002, and bathtub\_0074, respectively (from top to bottom).}
    \label{fig: mesh_quality1}
\end{figure*}

\begin{figure*}[htb]
    \centering
    \begin{subfigure}[b]{0.33\linewidth}
        \includegraphics[width=\linewidth, page=4]{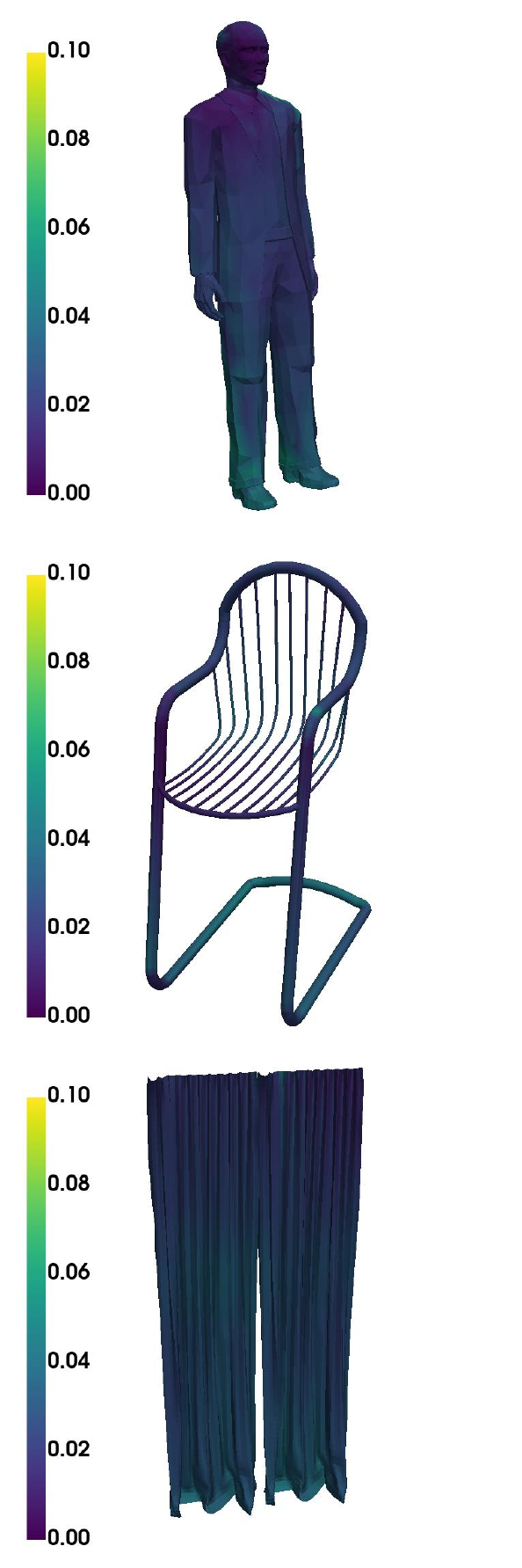}
        \caption{Visualization $L_1d$}
    \end{subfigure}
    \begin{subfigure}[b]{0.33\linewidth}
        \includegraphics[width=\linewidth, page=5]{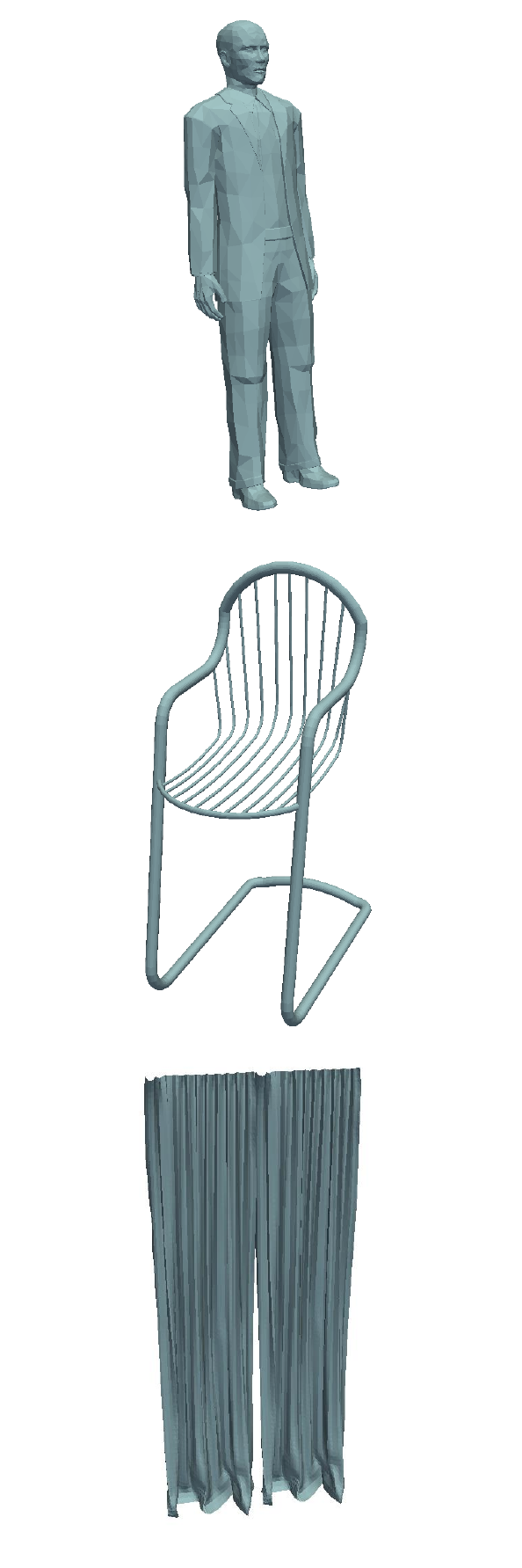}
        \caption{Original Mesh}
    \end{subfigure}
    \begin{subfigure}[b]{0.33\linewidth}
        \includegraphics[width=\linewidth, page=6]{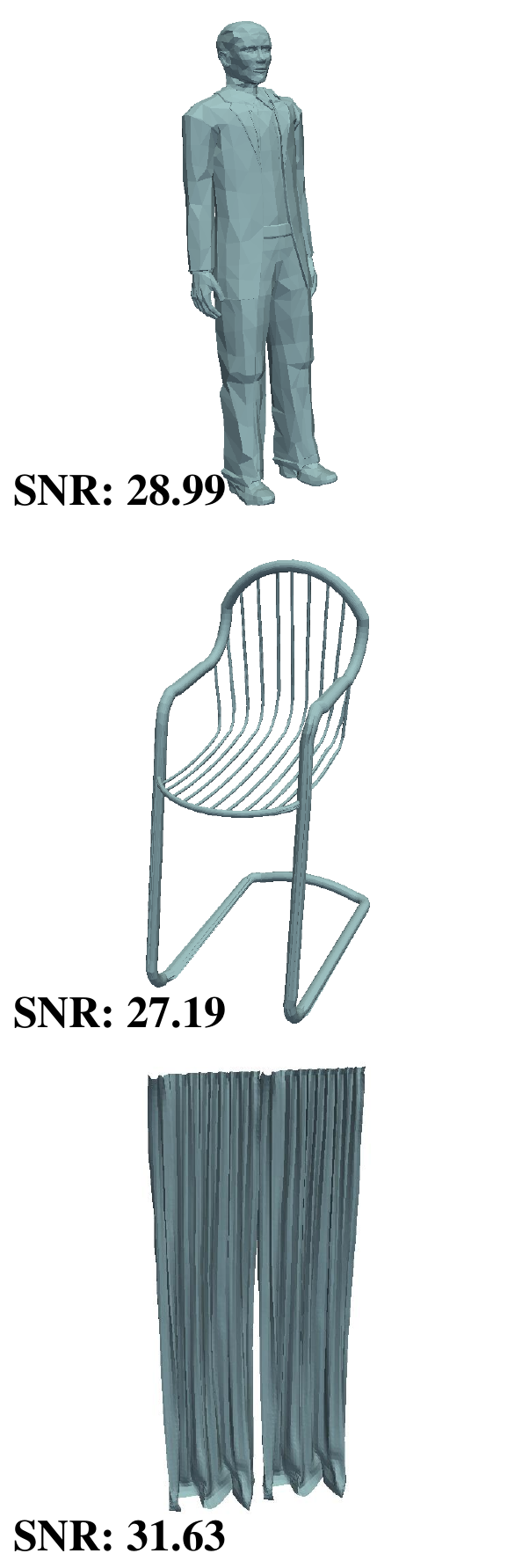}
        \caption{Watermarked Mesh}
    \end{subfigure}
    
    \caption{ModelNet40 Results to show the imperceptible watermarking: (a) L1 Norm of vertex difference ($L_1d$), (b) the original mesh, and (c) the watermarked mesh with \tool (ours). The visual data are person\_0045, chair\_0002, and curtain\_0011, respectively (from top to bottom).}
    \label{fig: mesh_quality2}
\end{figure*}

\begin{figure*}[htb]
    \centering
    \begin{subfigure}[b]{0.28\linewidth}
        \includegraphics[width=\linewidth, page=1]{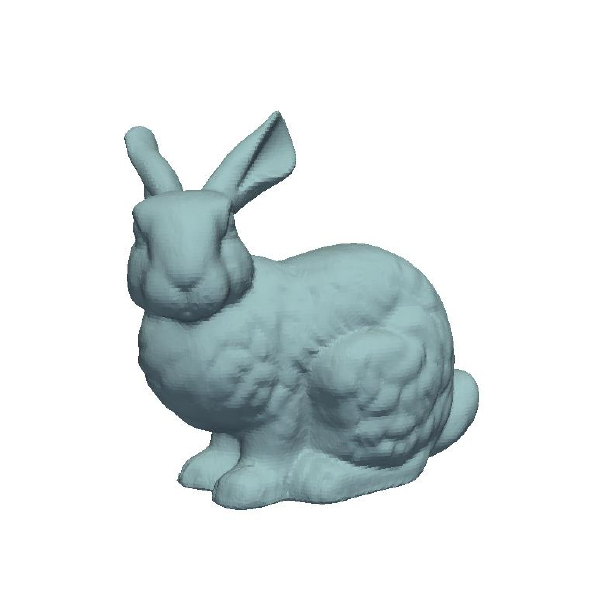}
        \caption{Original Mesh}
    \end{subfigure}
    \begin{subfigure}[b]{0.28\linewidth}
        \includegraphics[width=\linewidth, page=2]{figures/experiments/robustness/robustness_visual.pdf}
        \caption{Gauss($\sigma=0.002$)}
    \end{subfigure}
    \begin{subfigure}[b]{0.28\linewidth}
        \includegraphics[width=\linewidth, page=3]{figures/experiments/robustness/robustness_visual.pdf}
        \caption{Gauss($\sigma=0.01$)}
    \end{subfigure}
    \\
    \begin{subfigure}[b]{0.28\linewidth}
        \includegraphics[width=\linewidth, page=4]{figures/experiments/robustness/robustness_visual.pdf}
        \caption{Rotation($\alpha=0.2\pi$)}
    \end{subfigure}
    \begin{subfigure}[b]{0.28\linewidth}
        \includegraphics[width=\linewidth, page=5]{figures/experiments/robustness/robustness_visual.pdf}
        \caption{Crop($c=0.1$)}
    \end{subfigure}
    \begin{subfigure}[b]{0.28\linewidth}
        \includegraphics[width=\linewidth, page=6]{figures/experiments/robustness/robustness_visual.pdf}
        \caption{Implicit Laplacian Smooth ($\lambda_I=1.0$)}
    \end{subfigure}
    \\
    \begin{subfigure}[b]{0.28\linewidth}
        \includegraphics[width=\linewidth, page=7]{figures/experiments/robustness/robustness_visual.pdf}
        \caption{Implicit Laplacian Smooth ($\lambda_I=5.0$)}
    \end{subfigure}
    \begin{subfigure}[b]{0.28\linewidth}
        \includegraphics[width=\linewidth, page=8]{figures/experiments/robustness/robustness_visual.pdf}
        \caption{Implicit Laplacian Smooth ($\lambda_I=10$)}
    \end{subfigure}
    \begin{subfigure}[b]{0.28\linewidth}
        \includegraphics[width=\linewidth, page=9]{figures/experiments/robustness/robustness_visual.pdf}
        \caption{Draco compression ($N_q=15$)}
    \end{subfigure}
    \\
    \begin{subfigure}[b]{0.28\linewidth}
        \includegraphics[width=\linewidth, page=10]{figures/experiments/robustness/robustness_visual.pdf}
        \caption{Draco compression ($N_q=10$)}
    \end{subfigure}
    \begin{subfigure}[b]{0.28\linewidth}
        \includegraphics[width=\linewidth, page=11]{figures/experiments/robustness/robustness_visual.pdf}
        \caption{Draco compression ($N_q=5$)}
    \end{subfigure}
    
    \caption{To increase attack strength, we increase (b-c) Gaussian noise $\sigma$, (e) cropping $c$, (f-h) smoothing $\lambda_I$, and (i-k) Draco $N_q$. $N_q$ is the number of quantization bits to be compressed, and the rest of the bits are abandoned.}
    \label{fig: robustness_visual}
\end{figure*}

\begin{figure*}[htb]
    \centering
    \begin{subfigure}[b]{0.45\linewidth}
        \includegraphics[width=\linewidth, page=12]{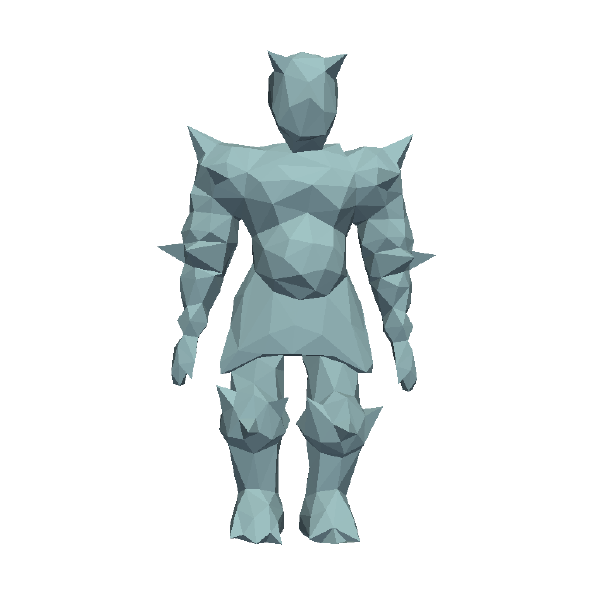}
        \caption{Original Mesh}
    \end{subfigure}
    \begin{subfigure}[b]{0.45\linewidth}
        \includegraphics[width=\linewidth, page=13]{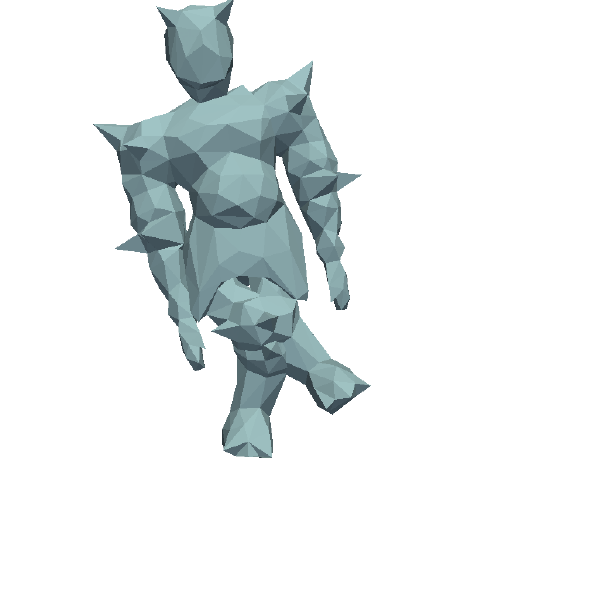}
        \caption{ARAP deformed}
    \end{subfigure}
    
    \caption{As rigid as possible (ARAP) deformation example.}
    \label{fig: arap}
\end{figure*}

All the coordinates of the visualized mesh are normalized to $[-1,1]$. For attacked mesh, we first normalize them, and then we apply arbitrary attacks. We first provide six more visualization results in Fig. \ref{fig: mesh_quality1} and Fig. \ref{fig: mesh_quality2}. We further visualize the mesh after attacks in Fig. \ref{fig: robustness_visual}. We also visualize the mesh after ARAP deformation in \ref{fig: arap}

